\documentclass[aip,jap,showpacs,preprintnumbers,amsmath,amssymb,superscriptaddress,floatfix,preprint]{revtex4-1}

\usepackage{hyperref}
\usepackage{graphicx}
\usepackage{color}
\usepackage{dcolumn}
\usepackage{bm}
\usepackage[latin1]{inputenc}
\usepackage{footnote}
\usepackage{epstopdf}
\usepackage{url}
\usepackage{balance}

\begin{document}

\title{Local electromagnetic properties and hysteresis losses in Non-Uniformly wound 2G-HTS Racetrack Coils}

\author{B. C. Robert}
\email[Electronic address: ]{bcr4@leicester.ac.uk}
\affiliation{Department of Engineering and Leicester Institute for Space $\&$ Earth Observation Science of the University of Leicester, Leicester, LE17RH, U.K.}

\author{M. U. Fareed}
\affiliation{Department of Engineering and Leicester Institute for Space $\&$ Earth Observation Science of the University of Leicester, Leicester, LE17RH, U.K.}

\author{H. S. Ruiz}
\email[Electronic address: ]{dr.harold.ruiz@leicester.ac.uk}
\affiliation{Department of Engineering and Leicester Institute for Space $\&$ Earth Observation Science of the University of Leicester, Leicester, LE17RH, U.K.}

\date{\today}

\begin{abstract}
A noteworthy physical dependence of the hysteresis losses with the axial winding misalignment of superconducting racetrack coils made with commercial \textcolor{black}{Second Generation High Temperature Superconducting (2G-HTS)} tapes is reported. A comprehensive study on the influence of the turn-to-turn misalignment factor on the local electromagnetic properties of individual turns, is presented by considering six different coil arrangements and ten amplitudes for the applied alternating transport current, $I_{a}$, together with an experimentally determined function for the magneto-angular anisotropy properties of the critical current density, $J_{c}(B,\theta)$, across the superconducting tape. It has been found that for moderate to low applied currents $I_{a} \leq 0.6~I_{c0}$, with $I_{c0}$ the self-field critical current of individual tapes, the resulting hysteretic losses under extreme winding deformations can lead to an increase in the energy losses of up to $25\%$ the losses generated by a perfectly wound coil. High level meshing considerations have been applied in order to get a realistic account of the local and global electromagnetic properties of racetrack coils, including a mapping of the flux front dynamics with well defined zones for the occurrence of magnetization currents, transport currents, and flux-free cores, what simultaneously has enabled an adequate resolution for determining the experimental conditions when turn-to-turn misalignments of the order of $20~\mu m$-$100~\mu m$ in a 20 turns $4~mm$ wide racetrack coil can lead not only to the increment of the AC losses but also to its reduction. In this sense, we shown that for transport current amplitudes $I_{a}>0.7~I_{c0}$, a slight reduction in the hysteresis losses can be achieved as a consequence of the winding displacement which is at the same time connected with the size reduction of the flux free core at the coil central turns. Our findings can be used as a practical benchmark to determine the relative losses for any 2G-HTS racetrack coil application, unveiling the physical fingerprints that possible coil winding misalignments could infer. 

\end{abstract}

\keywords{AC losses, hysteresis, superconducting coils, critical current density.}
%
\maketitle

\section{Introduction}\label{Sec.I}

High temperature superconducting coils with \textcolor{black}{Rare-Earth Barium-Copper-Oxide (ReBCO)} coated conductors are the cornerstone of nearly all current and envisaged high-power density applications taking advantage of the local electromagnetic properties of type-II superconductors~\cite{Barnes2004,Gieras2009,Kalsi2011,Rey2015,Marzi2016,Shen2018}. Attaining a clear understanding of the physical and designing parameters that might render to the minimization of AC losses in these systems is, by default, one of the most important subjects in the physics and engineering of applied superconductivity.\cite{Shen2018,Messina2016,Ainslie2015,ZhangAPL2012,Pardo2008,Clem2007SUST,Polak2006} 
However, despite many experimental and theoretical studies have been performed on the AC losses of HTS racetrack coils,\textcolor{black}{\cite{Martins2017,Wang2016,Pardo2013,Hong2011,Swaffield2014}} these always assume that the coil is perfectly wound, neglecting thence any influence of a possible misalignment between the coil turns (tapes), a situation that is likely to happen either during the coil manufacturing, their assembling in practical applications, or even due to possible axial alterations caused by extrinsic magnetic, thermal, or mechanical pressure over the coil turns. In fact, despite than in the last two decades the applied superconductivity field has experienced a significant increase in the physical understanding of the local electromagnetic properties of type II superconductors, including the second generation of high temperature superconducting (2G-HTS) tapes, it is the high aspect ratio of the 2G-HTS tapes what from the computational perspective is still imposing significant challenges to understand their performance in practical superconducting machines. This has left questions such as How does an unintentional misalignment in the winding of a superconducting coil impact its energy losses and magnetic field profile? It is in fact a negligible factor as it has been commonly assumed? or Can it actually deteriorate or improve the coil energy efficiency features? Motivated by these long standing questions and their practical importance from both the engineering and physics perspectives, below we present a unified numerical approach supported by high-level meshing conditions and a generalized description of experimentally measured material laws in 2G-HTS tapes~\cite{Ruiz2017SUST}, which focus more on the local understanding of the physical properties of racetrack coils rather than on commonly used mesh simplifications for reducing the expected computational burden. 

Thus, we present a comprehensive computational study of the profiles of current density, magnetic flux features, and AC losses of racetrack coils with 20 turns of commercially available 2G-HTS tapes (SCS4050),\cite{Ruiz2017SUST} either perfectly wound or with a maximum lateral misalignment factor $(\delta_{m})$ of $100~\mu m$ between adjacent turns, both subjected to AC transport currents of different amplitudes  (see Fig.\ref{Fig_1}). The numerical simulations are performed within the framework of the celebrated 2D H-formulation,\cite{Ruiz2018SciRep,Brambilla2007,Grilli2006,Hong2006} \textcolor{black}{implemented in COMSOL Multiphysics 5.3a, \cite{COMSOL2018}} together with what is up to now the most general material law for describing the $E$-$J$ properties of the 2G-HTS tapes at constant temperature, which extends the conventional Kim's critical state model to an experimentally validated magneto angular dependent $E$-$J$ power law~\cite{Ruiz2017SUST}. 

\section{Geometry considerations and modelling strategy}\label{Sec.II}

For the calculation of the AC losses, a total of sixty different cases have been studied, these considering up to ten different amplitudes of the alternating transport current, $I_{a}$, and six racetrack coils each with a distinctive misalignment factor $\delta_{m} = 20~\mu m$, $25~\mu m$, $30~\mu m$, $50~\mu m$, and  $100~\mu m$. \textcolor{black}{It is worth mentioning that the magnitude of the misalignment factors shown in this manuscript are to be understood in correlation with the overall displacement between the innermost and the outermost turn of the racetrack coil, also called the coil deformation, and therefore it needs to be read as a function of the width of the tape employed, $w_{s}$. In this sense, for a 20 turns $4~mm$ tape-width racetrack coil, these turn-to-turn misalignment factors correspond to coil deformations of $0.4~mm$ to $2~mm$, which for larger coils need to be understood as coil deformations in the range of $w_{s}/10$ to $w_{s}/2$. Certainly, as larger will be the coil as more likely will be the occurrence of these deformations, which might seem exaggerated within the size of the coil that has been considered here for numerical convergence and reasonable computational time. However, at this point is vital to emphasize that all the physical features encountered in this study, which for the sake of generality are presented in renormalized units, then will also be valid for racetrack coils with a greater number of turns. This is because the presented distribution of profiles of current density into our benchmark case of 20 turns, will allow to get a sufficiently large resolution for enabling a direct use of homogenization meshing techniques~\cite{Zermeno2013} or multi-scale models~\cite{Queval2016}, where a single one of our modelled superconducting coil turns could emulate the response of a whole stack of e.g., 50 turns out of a 1000 turns racetrack coil with turn-to-turn displacements as small as $0.4~\mu m$, where calculations of the AC losses are not expected to yield to errors larger than $2.5\%$. Moreover, it has to be noticed that although we have been performed simulations for smaller turn-to-turn displacements than $20~\mu m$, we have opted to show the results for displacement factors greater that this value as for $\delta_{m}<20~\mu m$ negligible differences $(<2\%)$ between the AC losses for perfectly wound coils and the misaligned coils have been obtained, which is in good agreement with the experimental results for gradient coil deformations in NMR microscopy, where deformations due to Lorentz's force can be as large as $1-10~\mu m$ depending upon the gradient strength and coil frame material~\cite{Chu1992}, although further considerations needs to be made regarding the magnetic field homogeneity as it will be shown in Sec.~\ref{Sec.III}.}


\begin{figure}
\begin{center}
{\includegraphics[width=0.88\textwidth]{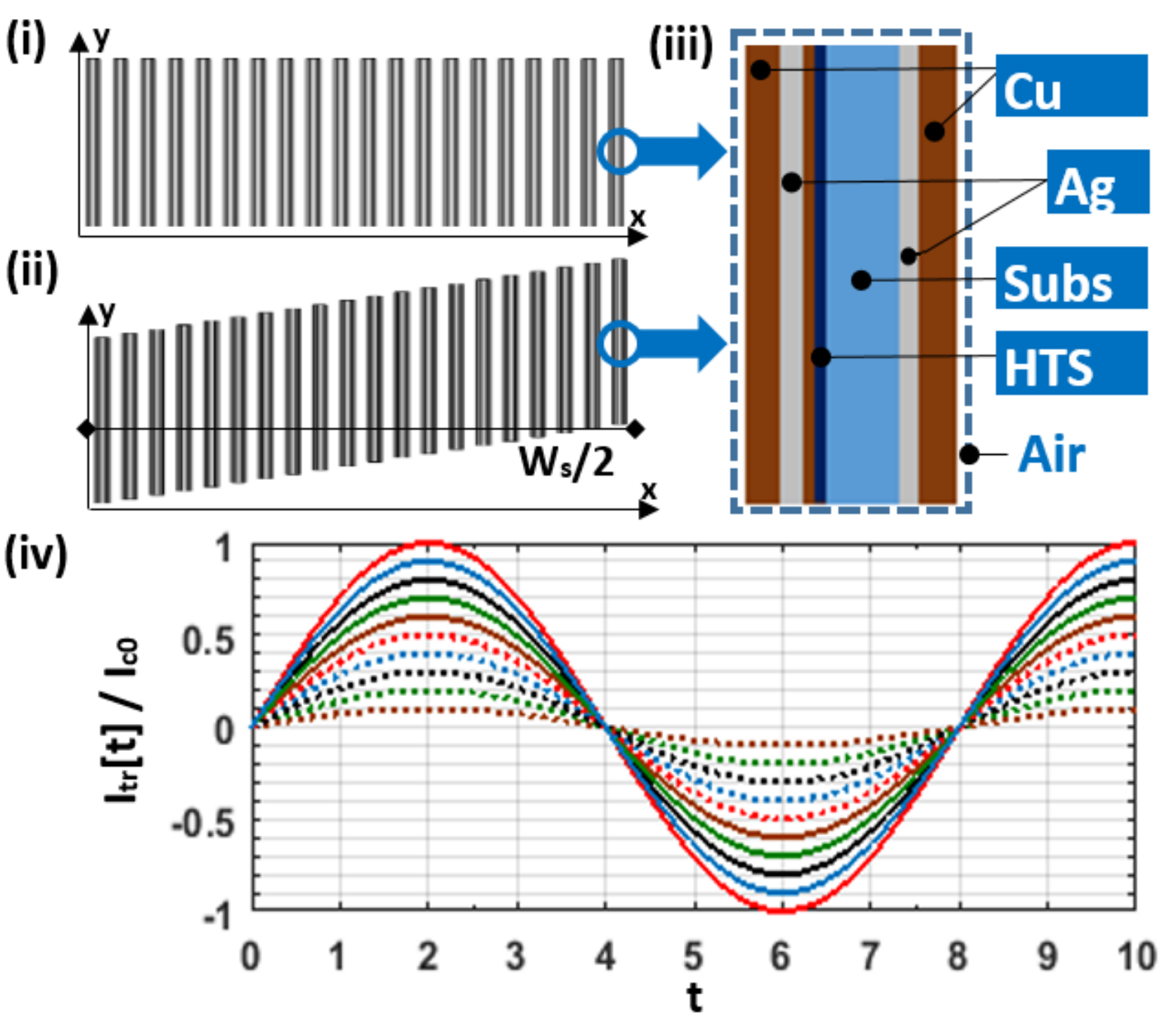}}
\caption{\label{Fig_1} Schematics of the cross-section of (i) a 20 turns perfectly aligned 2G-HTS racetrack coil, then (ii) subjected to a non-uniform winding $(\delta_{m} = 0.1~mm)$ with a $w_{s}/2$ misalignment between the innermost and outermost tapes. (iii) An individual 2G-HTS tape (not to scale) is shown for illustrative purposes, together with (iv) the time dynamics of the applied transport current with time steps labelled at each $\pi/4$ cycles.}
\end{center}
\end{figure}

Our study implies a significant computational endeavour which depends not only on the mesh quality of the physical domains, but also of the processors and RAM specs of the computer used for it. Thus, in order to overcome the computational difficulties imposed by the large aspect ratio of the SCS4050 2G-HTS tape (4~mm width, 0.055~mm thick), each individual coil turn has been modelled as an array of 6 tailored material domains [Fig.~\ref{Fig_1}.(iii)] each defined by 115 rectangular elements distributed along the width of the tape, times 4 layers for composing each one of the Cu coatings, 2 layers for the Ag protective layers, 4 for the HTS (ReBCO) domain, and finally 6 layers for the modelling of the \textcolor{black}{non-ferromagnetic} substrate. Then, a mapped triangular mesh with approximately 60000 elements has been implemented as a single ``air'' domain for modelling the insulating layers between the coil turns (0.2 mm thick) and the coil surrounding area ($1200~cm^{2}$), where a Dirichlet boundary condition $(H_{x}=0)$ is imposed at the center axis of the coil former ($9.4~cm$ width) taking advantage of the axial symmetry conditions of the problem for achieving minimal computing time. In fact, further symmetry conditions could be imposed for the perfectly wound coil, such that only its top half would need to be considered, reducing then the number of elements by $50\%$ or likewise, to double the number of elements for increasing the overall accuracy into our calculations. However, with the latter approach we have found that the mesh above defined for the entire width of the coil is sufficient to render the same accuracy level within the AC losses calculations. Moreover, given the fact that for misaligned windings dual axisymmetrical reductions cannot be imposed, it results compulsory to model the entire width of the coil in order to attain comparable results under the same physical and computational benchmark. In the light of this, it is worth mentioning that although further homogenization meshing techniques similar to the ones presented in~\onlinecite{Zermeno2013,Queval2016,Liang2017} could be applied to encapsulate the average behaviour of bundles of 2G-HTS tapes, i.e. by modelling multiple coil-turns into singular domains, therefore reducing the computing time, the inherent issue within this approach is the unavoidable loss of resolution and physical insight at a local level, where the patterns of transported or induced current within single turns results transposed into bulk-like descriptions. Thus, depending on the intensity of the applied current, the coil dimensions, and the size of the homogenized domains, these models can still give an accurate enough solution for the calculation of the AC losses in perfectly wound coils, but result inefficient for the modelling of misaligned coils, unless the number of coil turns is big enough such that the local distribution of profiles of current density within sequential turns is expected to show the same position for the flux front profiles, i.e., within the size of the smallest element where the flux front profile is to be located. Nevertheless, from the physical point of view where computing time reductions are not primordial, an accurate definition for the size of homogenized domains requires to have an ad hoc knowledge on the distribution of profiles of current density, being this one of the cornerstones of this paper.

From the physical point of view, the governing equations, Faraday's and Amp\`{e}re's law, are implemented within the magneto quasi-steady approach~\cite{Ruiz2009bPRB}, 

\begin{eqnarray}\label{Eq_1}
\nabla \times \bf{E} = - \mu_{0} \dfrac{\partial \bf{H}}{\partial t}  
\,,
\end{eqnarray}
\begin{eqnarray}\label{Eq_2}
\nabla \times \bf{H} = \bf{J}
\,,
\end{eqnarray}
with the $\textbf{E}-\textbf{J}$ material law of the SCS4050 2G-HTS tape accounting for the experimentally measured magnetoangular anysotropy in the critical current density,~\cite{Ruiz2017SUST} ${\bf J}_{c}(f({\bf B}), \theta)$, as follows, 
\begin{eqnarray}\label{Eq_3}
{\bf E} = \rho ({\bf J})\,: \rho = \dfrac{E_{c}}{J_{c}(f(B), \theta)}
\begin{vmatrix} \dfrac{\textbf{J}}{J_{c}(f(B), \theta)} \end{vmatrix}
^{n-1} \,,
\end{eqnarray}
with
\begin{eqnarray}\label{Eq_4}
J_{c}(f(B), \theta)=
\dfrac{J_{c0}}{\left(1\,+\,\epsilon_{\theta}\begin{vmatrix}\dfrac{\bf{B}}{B_{0}}\end{vmatrix}^{\alpha}\right)^{\beta}} \,,
\end{eqnarray}
\textcolor{black}{where $B_{0}$ and $\alpha$ are the so-called Kim's field parameters, $\beta$ is the flux creep exponent, and $\epsilon_{\theta} = \sqrt{\gamma^{-1} sin^{2}(\theta)+cos^{2}(\theta)}$ is known as the Blatter's angular anisotropy factor. The latter is a function of the electron mass anisotropy ratio $(\gamma)$ of the YBCO layer and, the angle of attack  $(\theta)$ of the magnetic field at the element of the SC mesh where the material law is being applied. Thus, $B_{0}$, $\alpha$, $\beta$, and $\gamma$ are constant parameters which depend only on the specific characteristics and fabrication of the  superconducting tape as being reported in Refs.~\onlinecite{Ruiz2017SUST,Zhang2018}, where for the specific case of the Superpower SCS4050 tape considered in this experiment, these parameters read as: $B_{0}=240~mT$, $\alpha=1$, $\beta=1.5$, and $\gamma=5.02$.}

From the technical point of view, one thousand time steps have been considered per $I_{tr}$ cycle, with computing times ranging from $\sim5$ hours for cases with $i_{a}=I_{a}/I_{c0}=0.1$, up to $\sim22$ hours for cases with $i_{a} = 1$, with the self-field critical current of the SCS4050 tape defined as $I_{c0}=114~A$ \cite{Ruiz2017SUST}. No significant computing time differences between the perfectly wound coil and the five designs for misaligned coils have been obtained, as long as the same current conditions are maintained. As 60 different simulations need to be ran, multiple CPUs have been used for this purpose. Nevertheless, we consider pertinent to call reader's attention on the fact that if a single CPU with the aforementioned specs is used, the study can take up to 31 days of full computing time, but this time can be even doubled in a more standard CPU of 3.0 GHz clock rate, 4~MB L2 cache, and 16~GB RAM memory. Also, it is worth mentioning that to the authors' knowledge most of the reported studies on superconducting racetrack coils~\cite{Brambilla2007,Ainslie2010,Hong2011,Zermeno2013} refer not only to perfectly wound magneto-isotropic coils, but for coils made of 12~mm wide tapes under particular transport current conditions aimed to assert specific experimental evidences. Therefore, none of the previous studies allow to infer a full physical description on the local time dynamics of flux front profiles inside of individual superconducting turns during the transient states between very low applied transport current (e.g., $I_{tr}=0.2~I_{c0}$) up to the threshold value of the critical current, being this one of the problems covered within the present study.


\begin{figure}
\begin{center}
{\includegraphics[width=0.88\textwidth]{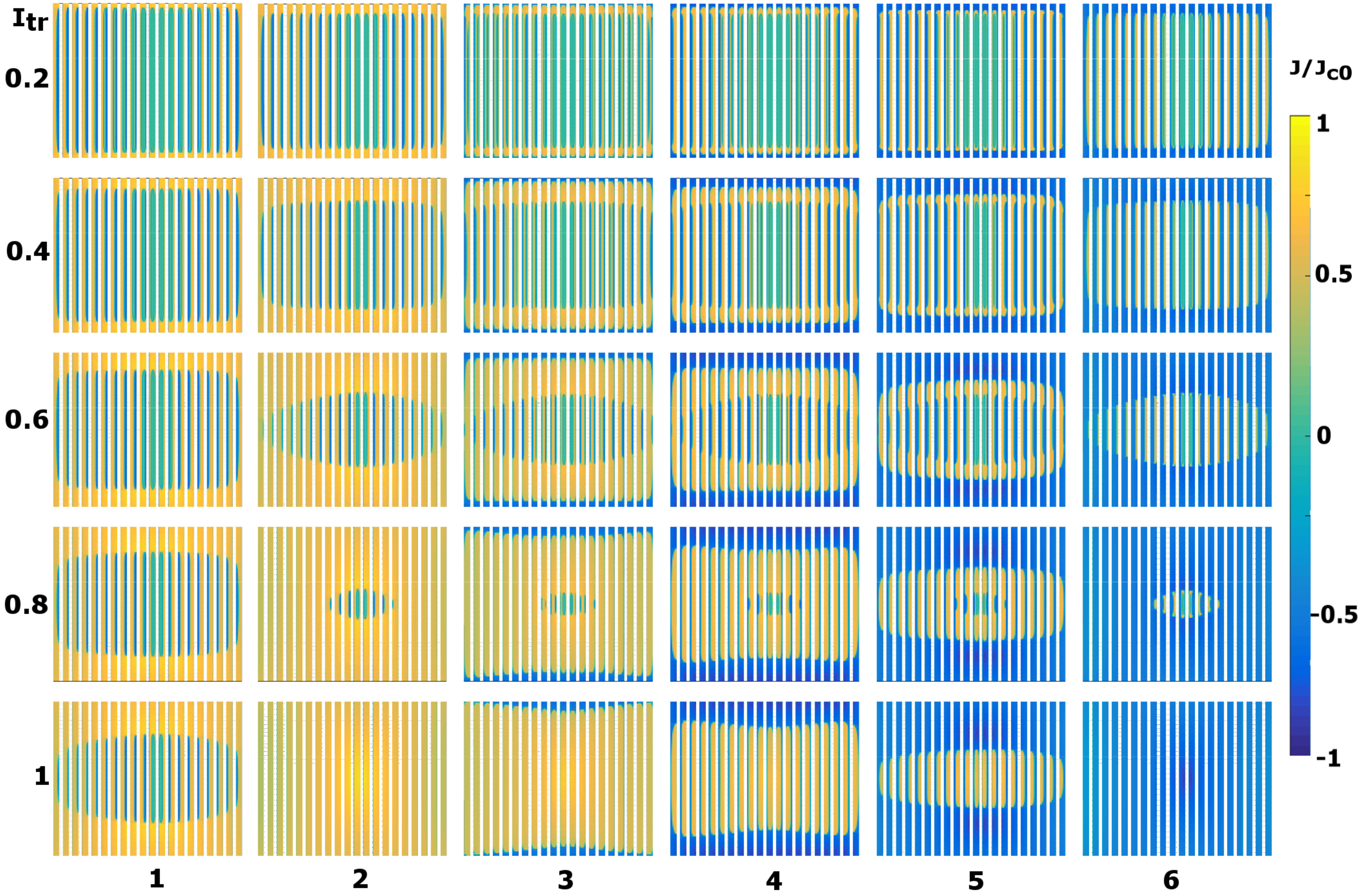}}
\caption{\label{Fig_2} Normalised profiles of current density $J/J_{c0}$ within the ReBCO layers (not to scale) of a 20 turns HTS racetrack coil in the configuration shown in Fig.~\ref{Fig_1} (i), and with applied AC currents of amplitude $i_{a}=I_{a}/I_{c0} = $0.2, \textcolor{black}{0.4}, 0.6, 0.8, and 1.0 from top to bottom, respectively. Time dynamics is presented as a sequence of labelled columns in accordance with the time steps displayed in Fig.~\ref{Fig_1} (iv).}
\end{center}
\end{figure}

\section{Electromagnetic characteristics of deformed winding racetrack coils}\label{Sec.III}

Below, a thorough analysis on the local electromagnetic characteristics of 2G-HTS racetrack coils governed by the time dynamics of the flux front profiles for transport and magnetization currents is presented, accounting for changes in macroscopical quantities such as the turn by turn critical current density with strong magneto-angular anisotropy, the resulting coil field and AC losses, in situations where slight misalignments in the coil winding could be present.

\subsection{Perfectly Aligned Windings (Case A)}

In Fig~\ref{Fig_2}, the normalized profiles of current density for 3/4 cycles of the applied current $(I_{tr}=I_{a}sin(\omega t))$ are presented, with amplitudes ranging from $0.2~I_{c0}$ up to $I_{c0}$, i.e., up to the threshold value for the self-field critical current measured on an individual HTS tape~\cite{Ruiz2017SUST}. For completing the full hysteretic picture of the time-dependent current dynamics shown in Fig.~\ref{Fig_1} (iv), it is just worth to mention that the profiles of current from the time-step 7 onwards are equivalent to the distribution of currents observed since the time-step 3 but with the profiles of current pointing in opposite directions, respectively. 

Within the magneto-quasi-steady approach~\cite{Ruiz2009bPRB}, the evolution of profiles of current density during the first monotonic ramp of the applied current (up to $I_{c0}$ at the time-step 2), can be extracted from the subplots displayed in the second column of Fig~\ref{Fig_2}, where it has to be noticed that the shape of the flux front enclosing the occurrence of magnetization currents, i.e., positive and negative currents flowing across the thickness of the HTS tapes (individual turns), follows the classical distribution of profiles of current density observed in HTS bulks of rectangular cross section~\cite{Badia2007}, with the exception that magnetization currents do not appear in the latter configuration. This result is only achieved if the coil turns are assumed as a system of electrically unconnected layers subjected to individual pointwise constraints for the injection of the transport current, as it is the case for stacks of 2G-HTS tapes,\cite{Ruiz2018SciRep,Ruiz2014APL} otherwise an oversized flux-free core will appear neglecting the influence of the magnetization currents on the individual turns of the coil. However, for low to moderate amplitudes, $i_{a}\leq 0.6$, the local distribution of current across the turns of the HTS coil shows a strong dependence on the self-induced magnetization currents with a flux free core (no current nor magnetic field) only visible within the central 2-3 turns. Moreover, we have observed that the full disappearance of the flux free core, i.e., the full penetration of the self-field to all coil turns is achieved at $i_{tr}\gtrsim 0.8$. Likewise, it has to be noticed that due to the consideration of the magneto-angular anisotropy of the SCS4050 tapes reported in Ref.~\cite{Ruiz2017SUST}, where the $J_{c}(B,\theta)$ function was experimentally measured, the intensity of the critical current across each one of the coil turns is strongly affected by the intensity of the self-field, such that at the innermost and outermost turns, it has been found that their average $I_{c}$ shows a detriment of up to $50\%$ the $I_{c0}$ value, decreasing radially towards the center of the wound tapes in a similar fashion to what has been reported in stacks of 2G-HTS tapes with simplified $J_{c}(B)$ models~\cite{Yuan2009,Yuan2010}. Another characteristic feature of the magneto angular dependence of $I_{c}$ that can be seen in the modelling of 2G-HTS coils for high intensities of the applied current  $(i_{tr} \geq 0.8)$, is the occurrence of a concave distortion of the flux front profiles enclosing the magnetization current domains which cannot be observed with simplified Kim-like $J_{c}(B)$ models~\cite{Zermeno2013,Prigozhin2011,Hong2011}.

\begin{figure}
\begin{center}
{\includegraphics[width=0.88\textwidth]{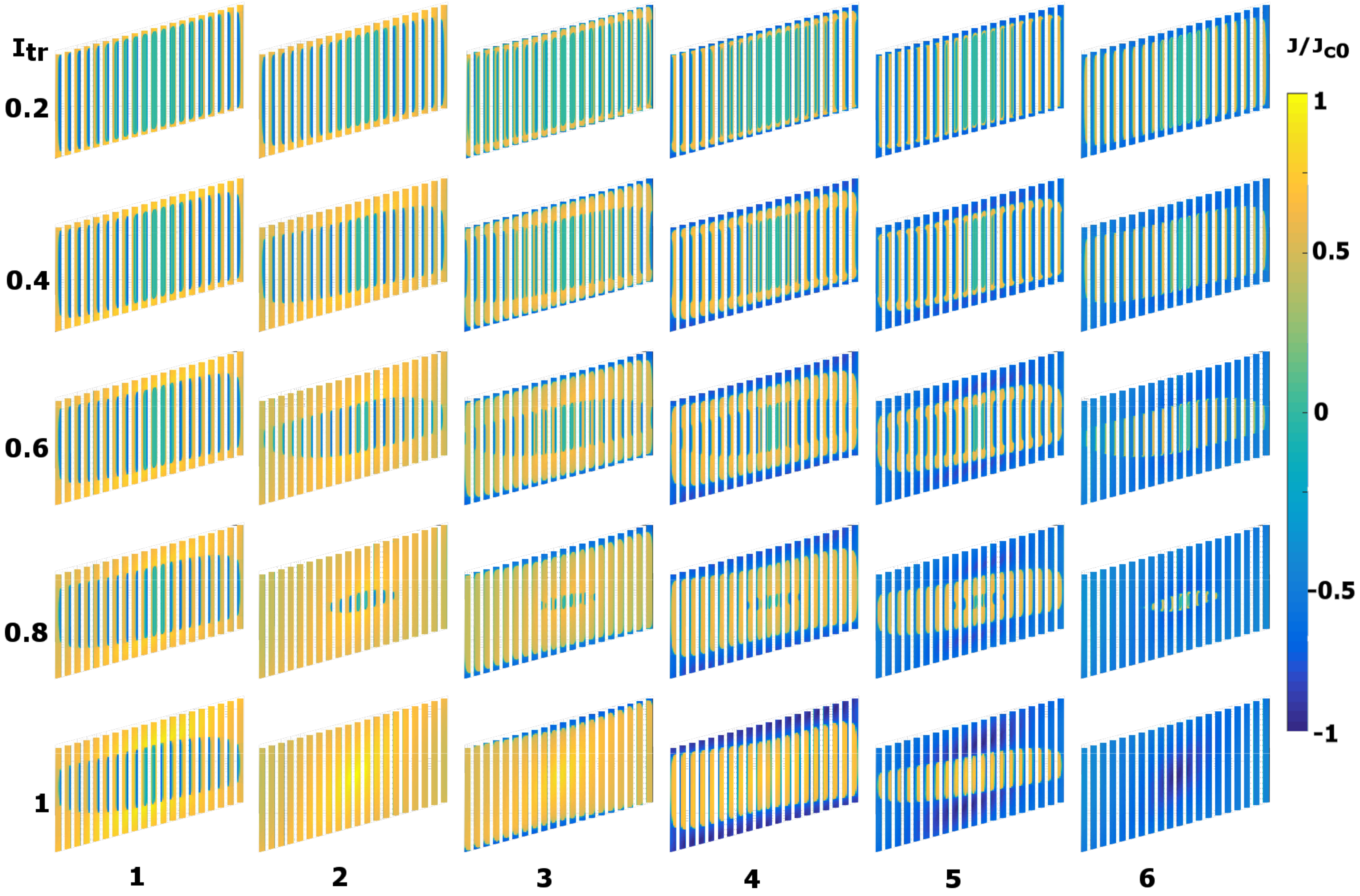}}
\caption{\label{Fig_3}Same as Fig~\ref{Fig_2} but for the misaligned coil shown in Fig.~\ref{Fig_1} (ii).
}
\end{center}
\end{figure}

\subsection{Non-Uniform Winding Distributions (Case B)}

Then, assuming that the coils turns are affected by a misalignment factor $\delta_{m} \times (w_{s}) \times (N-1)$, with $w_{s}$ defined as the width of the 2G-HTS tapes, and $N$ the number of coil turns, it has to be noticed that for a 20 turns SCS4050 coil, the factor $\delta_{m}=100~\mu m$ considers an extreme case where the outermost turn results displaced by half the width of the innermost tape $(2~mm)$ [Fig.~\ref{Fig_1} (ii)], which is the further case in this paper where the profiles of current density are being displayed (see Fig.~\ref{Fig_3}). In fact, although for a 20 turns coil the factor $\delta_{m}=100~\mu m$ results in a considerable large deformation of the racetrack coil, this condition still results of great relevance for coils with a much greater number of turns $(>100)$, where the maximum $\delta_{m}$ condition has been already established to be $\delta_{m} \leq 250~\mu m$~\cite{Duan2018}, as from the experimental point of view, turn by turn displacements greater than $250~\mu m$ would conduit to the delamination of the 2G-HTS tape if the applied applied magnetic stress level ($B^{2}/2\mu_{0}$) is superior to $5-MPa$~ \cite{Liu2016}, a situation that can be easily achieved in a superconducting coil operating at moderate to high transport currents. 

\begin{figure}
\begin{center}
{\includegraphics[width=0.88\textwidth]{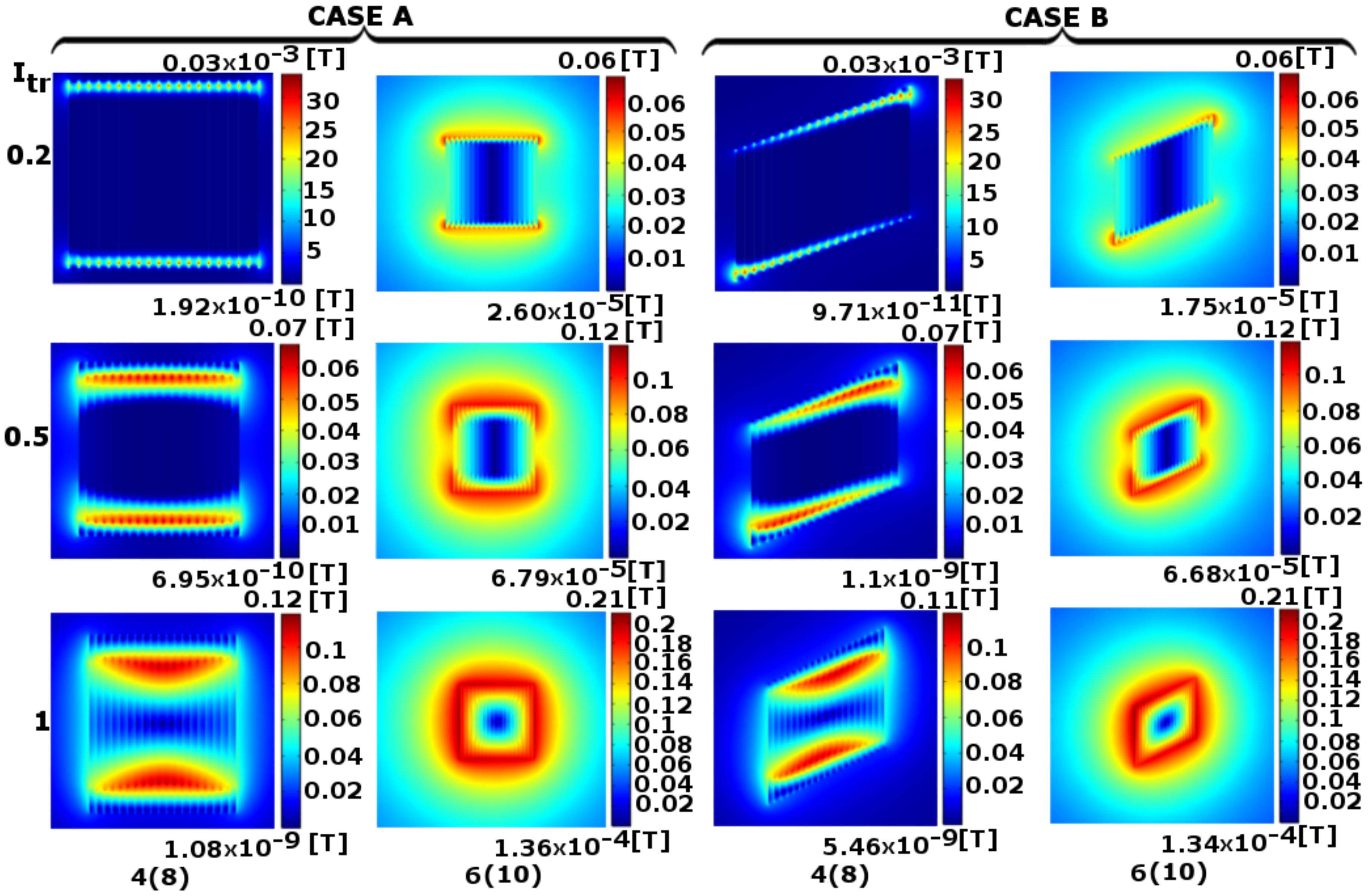}}
\caption{\label{Fig_4}Magnetic flux density $| B |$ created by the current density profiles displayed in (Case A) Fig~\ref{Fig_2} and (Case B) Fig~\ref{Fig_3} at the time steps 4 and 6 (or 8 and 10) displayed in Fig.~\ref{Fig_1}~(iv), with $i_{a}=0.2$ (top row), 0.5 (middle row), and 1 (bottom row). 
}
\end{center}
\end{figure}

In addition to our previous observations, in Fig.~\ref{Fig_3} it is to be noticed that there is a clear deflection of the flux front profiles due to the asymmetrical distribution between coil turns, which does not corresponds to a simple tilting of the original flux front profiles shown in Fig.~\ref{Fig_2}. In fact, the diversion of the flux front profile has been observed to be more acute towards the outermost and innermost coil turns, which means that local distribution of current density within each one of the turns is not longer symmetric between the upper and bottom half widths of the 2G-HTS tapes, implying that no symmetry reductions along the y-axis can be applied when winding deviations are present.  This phenomena is caused by the unbalance in the magnetic field experienced at the lateral edges of the tapes  (see Fig.~\ref{Fig_4}), where for instance the magnetic field at the lower corner of the innermost turn of the non-uniform coil $(\sim 129.65~mT)$ results lower than the magnetic field created by the uniformly wound coil $(\sim 140.87~mT)$ at the same position. Thus, this reduction in the local magnetic field at one of the lateral edges of the coil-turns implies an increment on the local intensity of the critical current density, what creates easier flow paths for the lines of transport current, pushing the magnetization currents towards the opposite edges in a similar way to what happens in isolated AC superconducting wires that are subject to synchronous and asynchronous magnetic field excitations~\cite{Ruiz2013JAP}. Consequently, the microscopic motion of the magnetization currents across each one of the individual turns, results reflected in a lateral inversion of the current density profiles across the innermost and outermost 2G-HTS tapes, what ultimately will result in a coherent deviation of the local density of power losses $\textbf{E}\times\textbf{J}$ that alters the hysteresis losses of the HTS coil. \textcolor{black}{Likewise, it is worth noticing that this change in the symmetry of the distribution of profiles of current density and the consequent unbalance on the magnetic field along the originally axisymmetric cut-lines of the perfectly wound coil, which has now suffered an axial deformation, could lead to up to a $8\%$ difference in the experimental readings for the magnetic field intensity and homogeneity, which can be of vital importance on the reliability figures for the readings of magnetic imaging applications~\cite{Moser2017} or high energy physics applications~\cite{Apollinari2015}.}

\subsection{AC losses Benchmark}

\begin{figure}
\begin{center}
{\includegraphics[width=0.88\textwidth]{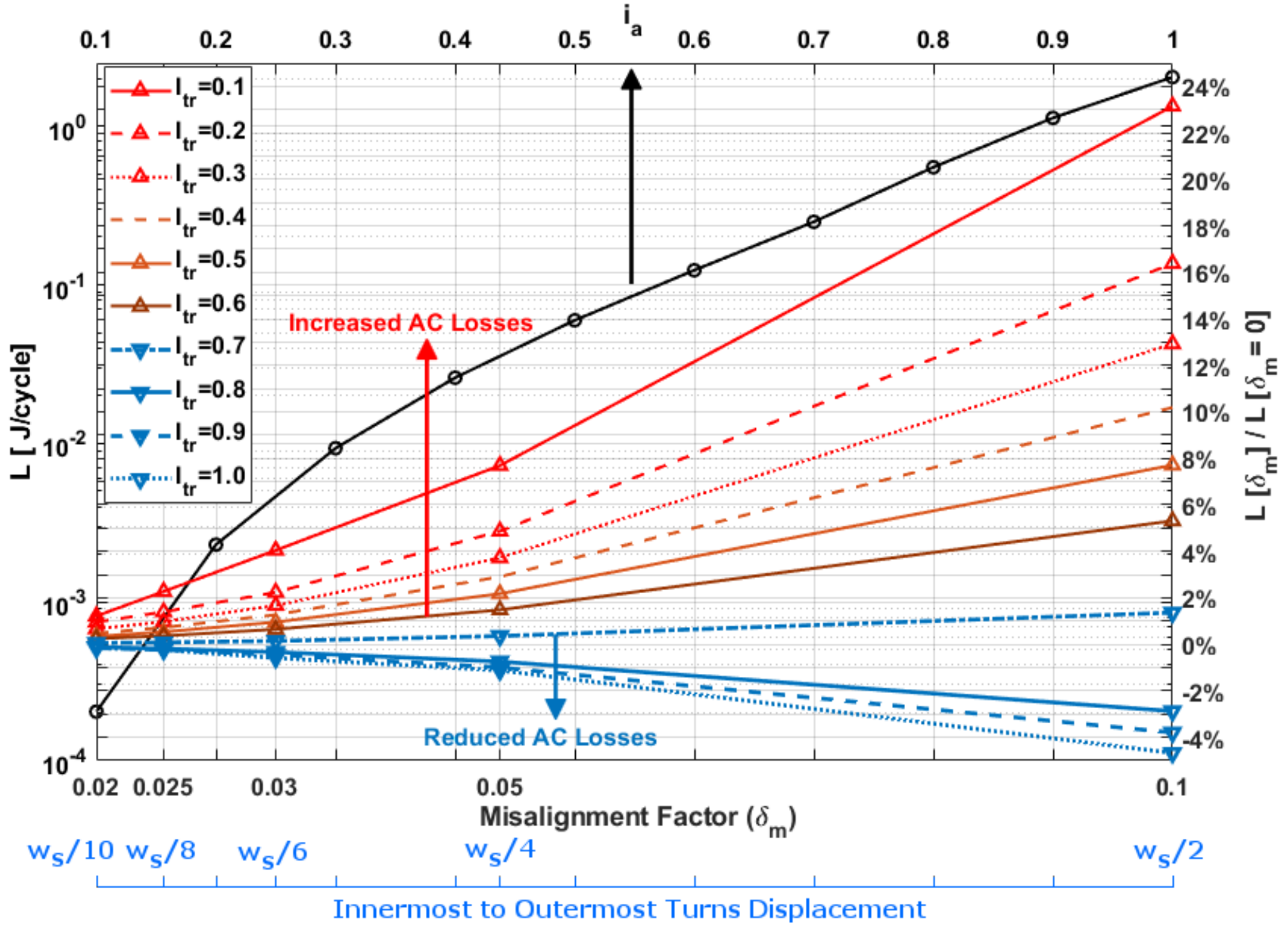}}
\caption{\label{Fig_5} \textcolor{black}{Top-left axes:} AC losses (Joules/cycle) for \textcolor{black}{the perfectly wound} HTS racetrack coil \textcolor{black}{with $\delta_{m}=0$}, as a function of the \textcolor{black}{normalized} transport current $I_{tr}=I_{a}/I_{c}=i_{a}$. \textcolor{black}{Bottom-right axes:} The AC losses curves \textcolor{black}{for the misaligned coils in percentage (renormalized) units, $L[\delta_{m}]/L[\delta_{m}-0]$,} are also depicted as a function of the individual misalignment factor $\delta_{m}$ for a 20 turns SCS4050 racetrack coil, \textcolor{black}{and in more general terms as a function of the coil deformation or} innermost-to-outermost turn displacement for a generic 2G-HTS tape of width $w_{s}$.
}
\end{center}
\end{figure}

It is commonly accepted that the average tolerance for the calculated AC losses in HTS coils by different electromagnetic formulations and material law models is about $10\%$, either when compared with experimental results or numerical predictions by competing computational models. Within this framework, the results presented within this section could be used by the community of researchers and companies interested in the deployment of superconducting technologies as a practical benchmark to determine the actual dependence of the AC losses on the axial alignment of non-uniform windings in HTS coils. In this sense, to avoid conjectural conclusions bonded to the coil dimensions, the AC losses values for the misaligned coils considered in this study are presented as a percentage function of the predicted AC losses for a perfectly wound coil. Therefore, in order to provide a clear insight on the effect of the misalignment factor on the AC losses of HTS racetrack coils, in Fig.~\ref{Fig_5} we present the calculated curve of AC losses for the perfectly wound coil as a function of the normalized amplitude of the applied transport current, $i_{a}$ (top curve, left/top axes), together with their percentage deviations for diverse turn-to-turn misalignment factors, $\delta_{m}$, and the corresponding displacements between the innermost and outermost turns of a generic HTS racetrack coil made of a superconducting tape of width $w_{s}$ (right/bottom axis).  

The results obtained for the misaligned coils are quite striking when compared with a perfectly wound coil, as strong increments in the AC losses can be observed for low amplitudes of the transport current, and even  compelling slight reductions of the hysteresis losses can be obtained for applied currents near to the critical threshold $I_{c0}$. Thus, within a general framework for the coil dimensions, we report that for displacements measured between the positioning of the innermost turn to the outermost turn of the HTS coil, the AC losses decreases as a function of the misalignment factor at currents $i_{a}\geq 0.8$, reaching up to a $5\%$ reduction at $i_{a}=1$ with $\delta_{m}=0.1$. However, for moderate to low currents, $i_{a}<0.8$, the magnitude of the AC losses rises monotonically up to an outstanding increase of nearly $24\%$ at $i_{a}=0.1$ and $\delta_{m}=0.1$.

These intriguing observations in the curves of AC losses can be understood with relation to the distribution of profiles of current density displayed in Fig.~\ref{Fig_3}, as therein is possible to witness how for applied currents $i_{a}\geq0.8$, there is no evidence of a flux-free core within the turns of the HTS coil, what implies that by increasing the magnitude of $i_{tr}$, the associated lines of transport current consume the remaining magnetization currents diminishing the hysteretic losses provided by the latter, a phenomenon that has been also evinced in type-II rounded superconducting cables within the critical state model~\cite{Ruiz2012APL}. On the other hand, the reported increase in the AC losses for misaligned coils is a result of the deformation of the flux front with asymmetric distributions of the profiles of current density and magnetic field across the different turns of the coil, which impacts in a greater measure the amount of magnetization currents that need to be diverted towards one of the edges of the HTS tapes, in the fashion explained in the previous section, as the $I_{a}$ reduces.

\section{Conclusions}\label{Sec.IV}

In this paper, the dependence of the AC losses with the axial alignment of windings in 2G HTS superconducting racetrack coils subjected to alternating currents is presented. As a study benchmark for the understanding of the local physical behaviour of macroscopic quantities such as the critical current density, magnetic field and hysteresis losses in racetrack coils, a 20 turns Superpower's SCS4050 has been modelled within a high level meshing approach which includes the magneto-angular dependence of the infield critical current density  ${\bf J}_{c} ({\bf B}, \theta)$, aiming to disclose the evolution of flux front profiles across the entire coil and the occurrence of magnetization and transport currents into the individual turns of the coil.

Based on the comprehensive computational study presented in this paper, we have demonstrated that besides the classical increment in several orders of magnitude in the AC losses of racetrack coils that can be observed as a function of the amplitude of the alternating transport current $(i_{a})$, a further increment of $5\%$ to $24\%$ over the hysteresis losses of perfectly wound coils $(\delta_{m}=0)$, can be obtained as consequence of possible deviations in the winding alignment for moderate $(i_{a}=0.6)$ to low $(i_{a}=0.1)$ alternating currents. On the other hand, we have found that for greater amplitudes of the transport current $(i_{a}\geq0.7)$ the change in the AC losses is either negligible at $i_{a}=0.7$, or even decreases in about $5\%$ of the losses calculated for a perfectly wound coil at $i_{a}=1$, i.e., when the applied transport current is as high as the the self-filed critical current $I_{c0}$ of the individual 2G-HTS tape.  Striking changes on the flux front profiles and local distributions of current density have been observed for the misaligned coils, and the increase or decrease on the AC losses has been found to be physically linked to either the local motion of magnetization currents or the disappearance of the flux free core, respectively. 

Although this is the first exhaustive computationally study on the influence of misaligned factors in the winding of 2G HTS racetrack coils, simultaneously, our results serve to validate some of the most intriguing electromagnetic features of perfectly wound coils which have been already predicted by integral and variational formulations such as the Minimum Magnetic Energy Variation Method~\cite{Souc2009,Pardo2008} and the $A-J$ formulation~\cite{Prigozhin2011} but with reduced meshing topologies, in particular, the fact that it is not possible to see radially oriented magnetization profiles across the HTS coil, it due to the disappearance of the radial field in the regions dominated by magnetization currents across the thickness of each coil turn. Similarly, we have observed that the distribution of averaged current density along the width of these tapes exhibit a slightly larger non-uniform distribution at the coil terminals when compared with the central turns of the coil, it due to the impact of the injected current causing a swift consumption of the magnetization currents across the innermost and outermost turns of the HTS coil. This phenomena has been initially predicted by Prigozhin and Sokolovsky in 2011~\cite{Prigozhin2011} on a coil of similar dimensions and equal number of turns, where it has been demonstrated that even within the Bean's critical state model~\cite{Bean1964}, $(|J| \leq |J_{c}|)$, this behaviour can occur in the innermost and outermost turn of the coil for $I_{a}=0.7~I_{c0}$. Additionally, we confirm under the full spectrum of current and magneto angular anisotropic conditions of real 2G-HTS tapes, one of the most important theoretical findings in \cite{Prigozhin2011}, where it has been demonstrated that when the $J_{c}$ dependence with the magnetic field is included through the classical Kim-like critical state model~\cite{Kim1964} for perpendicular and transverse components of the magnetic field in a racetrack coil with $I_{a}=0.7~I_{c0}$, the observed non-uniformity of $J_{c}$ is extended but limited to up to the second and penultimate turn of the HTS coil. In fact, given the increased resolution on our meshing and calculations, we show that the intriguing non-homogeneity of the current density over the width of the outer tapes is mainly caused by the magneto dependence of the critical current density, and in fact it does not only extends up to the second and penultimate turn of the coil, but at least up to the turn from which the elliptical-like shape flux front (encapsulating magnetization currents) starts to be  observed, a phenomena that can be observed in Figs.~\ref{Fig_2} \& \ref{Fig_3} at $I_{a}=0.8~I_{c0}$. Likewise, regardless of the missalignment factor, we confirm that at the outermost and innermost turns of a 2G-HTS racetrack coil, the averaged critical current density per turn experiences a detriment of about $50\%$ the self-field critical current density $J_{c0}$ when $i_{a} \geq 0.5$, being this in good agreement with previously reported studies in perfectly wound coils under simplified $J_{c}(B)$ Kim-like models\cite{Yuan2009,Yuan2010}. Thus, we expect our findings could be used as a practical benchmark for determining the source of possible deviations in the measured AC losses of racetrack coils that could be caused by manufacturing defects inherent to the winding process or, by mechanical deflections of the HTS tapes when the superconducting coil is positioned within a particular fixture.


\section*{Acknowledgement}

The authors acknowledge the use of the High Performance Computing Cluster Facilities (ALICE) provided by the University of Leicester. HSR has led the team and contributed with the analysis and writing of the paper. BCR thanks the  Niger Delta Development Commission for their funding support, and has contributed with the models development, analysis and plotting of results, and writing of the paper. MUF acknowledge the College of Science and Engineering Scholarship Unit of the University of Leicester, and has contributed through some of the computational tasks derived from this paper.

\section*{References}

\bibliography{References_Ruiz_Group_June_2019}

\end{document}